\documentclass[]{aa}  
\usepackage{xcolor}
\usepackage{graphicx}
\usepackage{txfonts}
\usepackage{lipsum}
\usepackage{subcaption}       
\usepackage{orcidlink}                                
\usepackage{lscape}            
                                
\usepackage{placeins}

\begin{document}

   \title{GeV emission in the region of Vela: a new view of the supernova remnant}

\author{Miguel Araya\orcidlink{0000-0002-0595-9267}\inst{\ref{inst1}} \and Santiago Ram\'irez\orcidlink{0009-0002-0323-7313}\inst{\ref{inst1}} \and Diego Bueso\inst{\ref{inst1}}\and Braulio J. Solano-Rojas\orcidlink{0000-0001-9912-0480}\inst{\ref{inst2}}}

\institute{Escuela de F\'isica, Universidad de Costa Rica, Montes de Oca, San Jos\'e, 11501-2060, Costa Rica, \email{miguel.araya@ucr.ac.cr}\label{inst1} \and ECCI/CITIC, Universidad de Costa Rica, Montes de Oca, San Jos\'e, 11501-2060, Costa Rica\label{inst2}
}

   \date{Received September XX, 2025}
 
  \abstract
   {The Vela supernova remnant (SNR), G$263.9-3.3$, and its pulsar wind nebula (PWN), Vela X, is one of the closest such systems, and it has been studied using observations across the electromagnetic spectrum. SNRs are known sources of gamma rays with energies from GeV to the TeV range. In the GeV band, a cluster of cataloged unidentified \emph{Fermi}-LAT point sources are found across the large angular extension of the Vela SNR.}
   {We aim to search for a high-energy signature associated to the SNR.}
   {We applied two independent machine learning algorithms to classify unidentified point sources in the Vela region by comparing their properties to those of known populations of \emph{Fermi} pulsars and active galactic nuclei. We analyzed LAT data and modeled the spectrum of any emission attributable to Vela using leptonic and hadronic processes typical of SNRs.}
   {We find that most of the ``point sources'' cataloged within the extent of Vela do not share characteristics with those of the two most common \emph{Fermi} point-like source populations and that even after the emission attributed to these ``point sources'' is subtracted, considerable residual emission is seen throughout Vela. Morphologically, most of the GeV emission is found within the shell of the SNR. We conclude that the majority of the cataloged point sources are likely spurious, and the GeV gamma rays come from an extended source, which we argue is the counterpart of the Vela SNR. Adopting a simple morphology given by a uniform disk for the emission the resulting extension is $\sim 6.5^\circ$. The northeastern portion of G$263.9-3.3$, where the ambient density is thought to be higher, is brighter in gamma rays than the south and west. The spectrum of the emission is best fit with a hadronic model. These facts make the hadronic origin for the gamma rays more likely.}
   {}


   \maketitle
   \nolinenumbers

\section{Introduction}
Supernova remnants (SNR) have been considered for many years plausible sources of Galactic cosmic rays. Particles gain energy and become relativistic in the shocks of SNRs, resulting in distinctive signatures such as synchrotron emission from radio to X-rays, and gamma-ray emission from inverse Compton scattering and non thermal Bremsstrahlung (leptonic origin) or neutral pion decay (hadronic origin). The Large Area Telescope (LAT) onboard the \emph{Fermi} satellite has been observing the sky since 2008, detecting photons with energies from $\sim 20$~MeV to more than 300~GeV \citep{2009ApJ...697.1071A}. The LAT fourth source catalog (4FGL) contains 7194 gamma-ray sources \citep{2020ApJS..247...33A,2023arXiv230712546B}, including a growing number of SNRs.

The $8^\circ$ diameter Vela SNR, G$263.9-3.3$, is one of the most studied and closest SNRs to Earth. It is a typical example of a middle-aged composite SNR, harboring the pulsar wind nebula (PWN) Vela X produced by the energetic Vela pulsar, PSR~B$0833-45$. This pulsar has a characteristic age $\tau_C = 11$~kyr, a spin period of $P = 89$~ms and a spin-down power $\dot{E} = 6.9 \times 10^{36}$~erg~s$^{-1}$. Its distance, $\sim 290$~pc, has been measured by parallax \citep{2001ApJ...561..930C,2003ApJ...596.1137D}. A shock radius of 20~pc, shock velocities of $660-1020$~km~s$^{-1}$ and an age of $(7-12)$~kyr have also been estimated \citep{2014A&A...561A.139S}.

Radio observations of Vela reveal a prominent $2^\circ \times 3^\circ$ region in the centre, dubbed Vela X. This is a synchrotron nebula likely produced by the pulsar \citep[e.g.,][]{1997ApJ...475..224F}. Another radio region seen within Vela, designated originally as Vela Z, is associated to a different SNR, Vela Junior (G$266.2-1.2$). \cite{2001A&A...372..636A} measured the integrated flux densities between 30 and 8400~MHz ($S_\nu$) of these two regions, as well as those of Vela Y, which corresponds to the northern shell of the Vela SNR. The radio spectral index of Vela Y is $\alpha \sim -0.70$ ($S_\nu \propto \nu^\alpha$). Using HI observations, \cite{1998AJ....116..813D} estimated an ambient atomic density of $1-2$~cm$^{-3}$ and an initial explosion kinetic energy of $(1-2.5)\times 10^{51}$~erg for the Vela SNR. They found evidence of atomic hydrogen located around the SNR, with denser emission in the north and east. \cite{2011A&A...525A.154S} explained the observed NE/SW asymmetry of the SNR. In their model, the wind from a nearby Wolf-Rayet star causes the NE part of the SNR to expand into a medium with an average density of shock-evaporated clouds that is about four times higher than that in the SW part (nucleon number densities 0.4 and 0.1~cm$^{-3}$, respectively). In this model, the required kinetic energy in the supernova explosion is $0.14\times 10^{51}$~erg. Regarding the high-energy particle content, \cite{2014A&A...561A.139S} proposed a scenario in which the multiple shocks seen inside the remnant lead to a uniform distribution of relativistic particles in the interior.

The Vela SNR is a bright source of X-rays \citep[see][and references therein]{2000A&A...362.1083L}. The NE appears brighter and less extended than the SW part of the SNR. \cite{1995Natur.373..587A} discovered several bow-shaped X-ray features outside the blast-wave front with the ROSAT telescope (the so-called Vela ``shrapnels''), which are likely dense ejecta fragments overtaking the main blast wave. Evidence for expansion in a clumpy medium has also been found in the NE, with clump densities of $0.5-1$~cm$^{-3}$ \citep{1999A&A...342..839B}. Embedded within the radio PWN, an X-ray feature known as the ``cocoon'' is seen \citep{Markwardt_1997}, likely produced by high-energy electrons from the pulsar. This elongated synchrotron structure has an extension of $\sim1.5^\circ$, with harder emission concentrated near the pulsar \citep{2018ApJ...865...86S}. More recently, \cite{2023A&A...676A..68M} performed a detailed study of the X-ray emission from Vela with the eROSITA telescope, revealing an SNR with a complex morphology and an extent of $10^\circ \times 8^\circ$. Interestingly, they discovered an extended synchrotron nebula around Vela X, elongated in the north-south direction with a diameter of $\sim 5^\circ$. This nebula could be produced by relatively recent injection from the pulsar after the ``crushing'' of the old nebula by the reverse shock, or by efficient difussion of high-energy particles from Vela X. Vela could thus be an example of a system where particles from the pulsar are trapped within the turbulent environment of the SNR, a possible explanation for the so-called TeV halos \citep{10.1093/mnras/stz1974}. Indeed, \cite{2018ApJ...866..143H} showed that the diffusion of electrons that have escaped from Vela X has to be suppressed given the flux of cosmic ray electrons seen on Earth.

Emission from Vela X has also been detected at gamma-ray energies. The \emph{Fermi}-LAT found the GeV counterpart to the radio nebula \citep{2010ApJ...713..146A,2013ApJ...774..110G} and H.E.S.S. found TeV emission in a region that encompasses the X-ray cocoon although it is somewhat more extended \citep{2006A&A...448L..43A}. The Vela pulsar itself is also a very bright GeV source \citep{2010ApJ...713..154A}. \cite{2018A&A...617A..78T} found that the LAT emission from Vela X is made of two components. Below 100~GeV the emission has a soft spectrum with spectral index $\sim 2.2$ and extends over a region consistent with the radio nebula. Above 100~GeV the spectral index is $\sim 0.9$ and this hard emission comes from a smaller region coinciding with the X-ray cocoon. The spectrum of the hard component connects smoothly to the one measured by H.E.S.S. at TeV energies. A more recent analysis by \cite{Lange_2025} including lower energy events and removing the pulsed emission from PSR~B$0833-45$ produced somewhat different results compared to those of \cite{2018A&A...617A..78T}. They concluded that two extended sources with similar spectral indices of $2.33$ constitute a more adequate description of Vela X. The larger source encompasses most of the known radio emission from Vela X and shows evidence of spectral hardening at the highest energies ($>100$~GeV), and the smaller source located to the NW of Vela X is only detected below 100~GeV. Given their similar spectral indices, the authors point out that the smaller source could be related to the PWN; however, an origin in the SNR or in an interaction of the PWN with SNR ejecta cannot be ruled out.

Regarding other GeV sources in the Vela region, there are 35 point sources without known association in the latest 4FGL catalog \citep[4FGL-DR4,][]{2022ApJS..260...53A,2023arXiv230712546B} located within $4.5^\circ$ of the coordinates RA,~Dec~$=(129.6^\circ, -45.3^\circ)$ near the centre of the SNR. Most of these sources are seen in the direction of the NE shell of Vela. We note that, within $8^\circ$ of these coordinates, there are five additional unassociated sources along the Galactic plane $\sim 3^\circ$ from the  southeastern shell of Vela, and four more at a similar angular distance to the south. A recent search for extended LAT sources in the Galactic plane found a candidate with a radius of $2.85^\circ$ (68\%-containment assuming a 2D Gaussian morphology) and a centre located within the extent of Vela X \citep{2024arXiv241107162A}. This source was potentially associated to the Vela pulsar, however, a cautious interpretation of this result was given in that work as the actual morphology of the source is likely not well represented by the assumed 2D Gaussian model. \cite{2024arXiv241107162A} also pointed out that a considerable 39\% change in the source extension TS (a measurement of the source significance, see below) was observed when adopting alternative models for the interstellar diffuse emission.

Unidentified LAT ``point sources'' that are clustered together could be related to a single object \citep{2025arXiv250403543C}. Further evidence to confirm this may come from detecting an extended source such as an SNR or a PWN in other bands of the electromagnetic spectrum \citep[e.g.,][]{2018ApJ...859...69A,2020MNRAS.492.5980A}. In this work we analyzed LAT data in Section \ref{data} to search for an extended GeV source in the region of Vela and to test its possible association to the SNR/PWN system. We adopted a classification procedure (explained in Section \ref{sourceclassification}) to decide which of the 4FGL point sources of unknown nature could in fact be part of Vela, and we present simple models that can explain the origin of the gamma rays in the SNR in Section \ref{discussion}.

\section{LAT data}\label{data}
We analyzed publicly available LAT data\footnote{See \url{https://fermi.gsfc.nasa.gov/ssc/data/} } collected between August 16, 2008 and June 20, 2025, corresponding to mission elapsed times (MET) 240623858~s to 772148810~s, in a $20^\circ \times 20^\circ$ region of interest (RoI) around the coordinates RA,~Dec~$=(129.6^\circ, -45.3^\circ)$ with the \texttt{fermitools}\footnote{See \url{https://github.com/fermi-lat/Fermitools-conda/}} version 2.5.1 through the \texttt{fermipy} package version 1.4.0 \citep{2017ICRC...35..824W}. Since the Vela pulsar is a very bright source of gamma rays we used an updated pulsar ephemeris solution\footnote{Kindly provided to us by M. Kerr.} valid in the same time range to assign pulse phases to the events using the \textit{Tempo2} software \citep{tempo2}. We defined the off-pulse phase window intervals $0-0.1$ and $0.7-1$ and removed events in the on-pulse window for the entire analysis. We set the \texttt{fermipy} parameter \texttt{expscale=0.4} which corrects the exposure due to our phase selection. We selected events with energies above 1~GeV for improved angular resolution to carry out a morphological study, and used the energy range $0.1-100$~GeV to extract the source spectrum. Since the PWN Vela X shows an additional high-energy component that is detected with the LAT above 100~GeV, we set this value as the upper energy range. For the spectral and morphological analyses we selected front and back-converted \texttt{SOURCE} class events (with the parameters \texttt{evclass}=128, \texttt{evtype}=3). To avoid contamination from the Earth's limb, the  maximum zenith angle was set to $105^\circ$ for the morphological study using events with energies above 1~GeV, and to $90^\circ$ for the spectral analysis in the entire energy range. We binned the data with a spatial scale of $0.05^\circ$ per pixel and ten bins per decade in energy for exposure calculations. We used the response functions appropriate for this data set, \texttt{P8R3\_SOURCE\_V3} and applied the maximum likelihood technique \citep{1996ApJ...461..396M} to model the data. The model of a source is convolved with the response functions to predict the number of counts in the spatial and energy bins. A fit of the free parameters maximizes the probability for the model to explain the data in each bin. The detection significance of a new source having one additional free parameter, for example, can be calculated as the square root of the test statistic (TS), defined as TS$=-2 \,\mbox{log}\,(\mathcal{L}_0/\mathcal{L})$, with $\mathcal{L}$ and $\mathcal{L}_0$ the maximum likelihood functions for a model with the new source and for the model without this additional source (the null hypothesis) respectively. Our starting model for the gamma-ray emission in the region included the sources found in the 4FGL-DR4 catalog. The diffuse Galactic emission is described by the file \texttt{gll\_iem\_v07.fits} and the isotropic emission and residual cosmic-ray background is given by \texttt{iso\_P8R3\_SOURCE\_V3\_v1.txt}, both provided with the \texttt{fermitools}. The energy dispersion correction was applied to all sources except for the isotropic component.

\subsection{Previously studied LAT sources in the region}\label{latsources}
From the unassociated sources in the region of Vela, we considered the source 4FGL~J$0822.8-4207$ to be plausibly associated with the SNR Puppis A \citep{2022MNRAS.510.2277A,2025A&A...701A.206G} and replaced its point-like model with a small disk as found by \cite{2025A&A...701A.206G}. The source PS~J$0824.0-4329$, located to the south of Puppis A and proposed by \cite{2025A&A...701A.206G} to better account for the emission in this region, was included in the model. Furthermore, these authors used a different template to describe the emission from Puppis A compared to the one in the catalog, which is not sufficiently adequate to explain the emission to the west of Puppis A. Therefore, we included two additional point sources in the model to account for this emission and optimized their locations with our data, resulting in the coordinates (RA, Dec) = ($125.07^\circ$, $-42.89^\circ$) and (RA, Dec) = ($125.46^\circ$, $-43.21^\circ$). Additionally, we adopted the model obtained by \cite{Lange_2025} to describe the emission from Vela X with two geometrical templates consisting of a Gaussian and a disk to replace the source 4FGL~J$0834.3-4542$e. Finally, according to \cite{2024NatAs...8..530P}, several stellar clusters in the Vela molecular cloud ridge are sources of gamma rays. These sources correspond to the unassociated sources 4FGL~J$0844.9-4117$, 4FGL~J$0859.3-4342$, 4FGL~J$0900.2-4608$, 4FGL~J$0857.7-4256$c, which we kept in the model. We also adopted their extended sources to model the emission from RCW 27 (replacing the point source 4FGL~J$0838.4-3952$, located $\sim 5.4^\circ$ away from the centre of the RoI), RCW 38 \citep[replacing 4FGL~J$0859.2-4729$, see also][]{2024MNRAS.530.1144G} and an additional source, which they labeled `gas core', replacing 4FGL~J$0900.5-4434$c and 4FGL~J$0901.1-4456$c. Since G$264.681+00.272$ and G$264.220+00.216$ are classified as candidate HII regions \citep{2014ApJS..212....1A} we did not consider their association to 4FGL~J$0853.1-4407$ as firmly established. This leaves 26 point sources of unknown nature within $4.5^\circ$ from the centre of the analysis region, most of which are located within the extent of the Vela SNR shell. We found that even after subtracting the emission from these sources, considerable residual emission is seen across the shell of Vela, but a more adequate study of the SNR features at GeV energies requires deciding which unidentified point sources could be related to it. In order to know which of these sources could potentially be associated to independent objects, and thus should be kept in the model, we performed two classification schemes described in the following section.

\subsection{Source classification}\label{sourceclassification}
In the original 4FGL release \citep{2020ApJS..247...33A} most of the identified or associated point sources ($\sim 63$\%) correspond to blazars, an extreme type of active galactic nuclei (AGN). Based on their optical properties, these AGN can be classified into BL Lacertae objects (BLLs) and flat spectrum radio quasars (FSRQs). The second most common class of identified or associated point sources corresponds to pulsars. AGN and pulsars show distinct spectral features in the 4FGL catalog \citep{2020ApJS..247...33A}. Around $27$\% of the cataloged point sources do not have a known counterpart at other wavelengths.

Machine learning provides adequate classification and characterization of point-like LAT sources \citep[see, e.g.,][]{2016ApJ...820....8S,XIAO2020100387,CHIARO202140,2021MNRAS.507.4061F,2022A&A...660A..87B,2023MNRAS.521.6195M,stad2813}. We adopted two pipelines in order to classify sources of unknown nature based on their gamma-ray properties: (i) a multiclass neural network\footnote{Architecture and training follow standard deep-learning practice for multi-class classification according to \url{https://keras.io}} that decides whether a source is an AGN or not, and (ii) a calibrated support vector machine (SVM) ensemble that separates pulsar-like from non–pulsar-like sources to flag pulsar candidates. Both models were trained on the spectral and variability parameters of the 4FGL-DR4 catalog and return a probability that a given source belongs to any of the classes. For training, we used sources with a designation in the 4FGL \texttt{CLASS1} column (meaning that a source is identified or likely to belong to a certain class of objects). Unassociated 4FGL sources with an empty \texttt{CLASS1} entry were treated as unidentified objects to be classified and were not used as labelled training examples. Finally, we cross-checked the results obtained in methods (i) and (ii) for consistency.

\subsubsection{AGN candidate selection}\label{agnClass}
For the AGN classifier we only kept sources with \texttt{CLASS1} = BLL or FSRQ as positive examples, and we grouped all other known classes that are not AGN (pulsars of any type, pulsar wind nebulae, supernova remnants, globular clusters, star-forming regions, X-ray/$\gamma$-ray binaries, novae, normal or starburst galaxies) into a single \texttt{nonAGN} label. Blazar candidates of uncertain type (BCU) and AGN subclasses that are neither BLL nor FSRQ were excluded from the training set and are only used later as independent test samples. 4FGL AGN subclasses other than BLL, FSRQ and BCU constitute less than $\sim$2\% of AGN and can be neglected. This choice follows common LAT source classification practice where blazars dominate the AGN population and rare subclasses are not modeled as standalone classes \citep[e.g.,][]{kovaevi_2020_classification,2023MNRAS.525.1731C}.

Our multiclass supervised neural classifier thus assigns LAT point-like sources to the classes \texttt{BLL}, \texttt{FSRQ} or \texttt{nonAGN}. We handled class imbalance with SMOTENC. We labeled as AGN those with high posterior probabilities $P(\mathrm{BLL})$ and $P(\mathrm{FSRQ})$. We chose as training parameters 15 descriptors in the 4FGL catalog that encode spectral shape (i.e., describe the function corresponding to the differential energy spectrum, $\frac{dN}{dE}$, where $E$ is the energy) and variability, such as spectral indices for three common LAT spectral models, the power-law (PL), the log-parabola (LP), and the power-law with exponential cutoff (PLEC), together with the curvature parameters \texttt{LP\_Beta}, \texttt{PLEC\_ExpfactorS} and \texttt{PLEC\_Exp\_Index}, which quantify the deviation from a pure power-law, curvature significance (i.e., \texttt{LP\_SigCurv}, \texttt{PLEC\_SigCurv}), flux density at the pivot energy for each model, predicted photon yield ($N_{\mathrm{pred}}$), and variability indicators (the variability index and the fractional variability). To make the categorical field \texttt{SpectrumType} usable by the classifier we created three binary features associated to the spectral shapes PL, LP and PLEC, setting the one that matches the source's spectral model to 1 and the others to 0. It is advantageous to expose the network to several mathematically related descriptors even if some of the parameters are strongly correlated. During exploratory data analysis we decided to remove catalog entries associated to the ``peak energy'' (the energy for which $E^2\frac{dN}{dE}$ shows a maximum), which had substantial missing entries, such as \texttt{LP\_EPeak}, \texttt{PLEC\_EPeak}, and \texttt{Flux\_Peak}. Retaining them would either force heavy imputation or shrink the effective sample size without adding discriminative power beyond the curvature proxies we kept (\texttt{LP\_SigCurv}, \texttt{PLEC\_SigCurv}). We also excluded sources with undefined entries in the catalog values.

We used 80\% of the sample for training and a Keras/TensorFlow multilayer perceptron \citep[MLP,][]{abadi2016tensorflow,gron_2019_handson} tuned with \texttt{Optuna} \citep{akiba2019optuna}. Stratified shuffle 10-fold cross-validation\footnote{As implemented in \texttt{scikit-learn} (e.g., \texttt{StratifiedShuffleSplit}/\texttt{StratifiedKFold}).} \citep{Kohavi1995,Pedregosa2011} yields an average accuracy of $85.4\%\pm1.4\%$ and a weighted F1 score \citep[defined as the harmonic mean of precision and recall,][]{sokolova2009systematic} of $0.854\pm0.014$, with per-class AUCs $>0.94$. Our final model selected from 100 independent trainings achieved $87.8\%$ accuracy and a weighted F1 score of $0.8775$, indicating stable generalization across the three classes. As an aside application, we used our trained neural network to classify the 4FGL set of blazar candidates of uncertain type (BCUs), resulting as follows: 828 were assigned to the class \texttt{BLL}, 566 to \texttt{FSRQ} and 229 to our class \texttt{nonAGN}, out of a total of 1,623 BCUs. The resulting fractions of BLL and FSRQs are comparable to those found by \cite{2023MNRAS.525.1731C}, who tested several classification algorithms.

\subsubsection{Pulsar candidate selection}
An SVM is a supervised method that learns a decision boundary that best separates two classes by maximizing the margin between them \citep{CortesVapnik1995}. We used the \texttt{LIBSV} solver via Python interfaces in \texttt{scikit-learn}\footnote{See \url{https://scikit-learn.org/}} and performed automated hyperparameter search with \texttt{Optuna}. \texttt{LIBSVM} is a widely used open-source library that implements SVM classification and regression with efficient, well-tested optimization routines and optional probability estimates \citep{ChangLin2011}.

A preprocessing of the data was performed to select the optimal subset of attributes that allows for accurate classification. First, the proportion of missing data in each attribute was studied, eliminating those with a very high percentage of missing values. To avoid the Hughes phenomenon (or curse of dimensionality) during training, two feature selection techniques were employed. First a Support Vector Machine Recursive Feature Elimination was applied to quantify the discriminative importance of each attribute, based on the procedures of \cite{Hughes}, and then a selection based on Pearson correlation was used to eliminate redundant attributes (pairs of attributes with high correlation between them). In this way, computational resources are optimized and the model's discriminative capacity improved.

For training, we selected labeled examples from the LAT catalog: 4067 point sources with firm non-pulsar associations (negative class) and 320 identified pulsars (positive class). Among the catalog attributes, \texttt{PLEC\_SigCurv} and the variability index were most informative, consistent with prior LAT results where pulsars show considerable spectral curvature and low variability, while many non-pulsars (e.g., AGN) are highly variable \citep{2022ApJS..260...53A}.

We divided the dataset into training and test sets, with 80\% of the data allocated to training to ensure that the models learn as much as possible from the available samples. Importantly, the original proportion of pulsar and non-pulsar sources in the catalog was preserved in both the training and test sets. Because pulsars are much more rare than non-pulsars (class imbalance), four training variants were constructed using sampling techniques: (i) the original set (unbalanced), (ii) oversampling (adds new minority-class examples by interpolating near existing ones; this densifies pulsar-like regions), (iii) undersampling/cleaning (removes ambiguous points where classes overlap, reducing noise near the decision boundary), and (iv) a combined oversampling+cleaning variant. One SVM was trained on each variant, and their outputs were combined into a weighted ensemble (models with stronger validation performance received higher weight). Model settings were tuned to prioritize accurate separation while controlling overfitting, using the pulsar class F1 score as the optimization target (as indicated earlier, this is the harmonic mean of precision and recall, where precision is the fraction of predicted pulsars that are truly pulsars, and recall or sensitivity, is the fraction of all true pulsars that are correctly recovered). It thus summarizes the trade-off between selecting clean samples and finding as many true pulsars as possible when the positive class is rare.

On the held-out test set, the ensemble achieved an F1 score of 0.89, improving over individual models and especially boosting recall (sensitivity) to 0.91. As SVM algorithms are deterministic, calibrated probabilities were obtained from the ensemble scores via Isotonic calibration \citep{zadrozny2002transforming}. Isotonic calibration is a non-parametric method that uses isotonic regression to fit a monotonically increasing piecewise constant function, mapping classifier scores to well-calibrated probabilities without assuming a specific functional form; when calibration is good, a predicted probability $p$ can be interpreted as an empirical frequency: among sources assigned probability $p$, roughly a fraction $\approx p$ are expected to be true pulsars in data with similar characteristics.

\subsubsection{Results of the classification}
A threshold of $80\%$ on the ensemble score was used to decide which sources are labeled as pulsar candidates, while a value of $70\%$ was used for AGN candidates (meaning $P(\mathrm{BLL})\geq 0.70$ or $P(\mathrm{FSRQ}) \geq 0.70$). These choices are supported by the evaluation metrics obtained when applying the trained models to the test sets: the resulting performance, observed through metrics such as precision, recall, and F1-score, confirms that these thresholds provide a reliable separation between classes when compared with the known classifications. Table \ref{tab1a} contains the results of the classification. No sources were classified as belonging to both the AGN and PSR groups, showing consistency between the independent algorithms. Only four sources were preliminarily assigned to the pulsar type (the sources in bold text in the table) while no sources were classified as AGN. Of the pulsar candidates, 4FGL~J$0854.8-4504$ showed a high pulsar probability (0.97). Using machine learning techniques \cite{2016ApJ...820....8S} also classified this source as a pulsar, however, dedicated X-ray observations by Chandra did not find a counterpart \citep{2024ApJ...961...26R}. If a classified source showed TS < 25 in the analysis in Section \ref{spectrum}, we removed it from the model (this was the case for 4FGL~J$0843.9-4224$c, see below).

\begin{table}[ht]
\caption{Source classification probabilities}
\resizebox{1\columnwidth}{!}{
\begin{tabular}{lccccc}
\hline\hline
\multicolumn{1}{c}{\textbf{Source}} &
  \multicolumn{2}{c}{\textbf{ANN}} &
  \multicolumn{1}{c}{\textbf{SVM ensemble}} \\ 
& 
  \multicolumn{1}{c}{\textbf{$P(\mbox{\texttt{BLL}})$}} &
  \multicolumn{1}{c}{\textbf{$P(\mbox{\texttt{FSRQ}})$}} &
  \textbf{  $P(\mbox{\texttt{PSR}})$  } \\ \hline
J$0828.4-4444$ & \multicolumn{1}{c}{0.00035} & \multicolumn{1}{c}{0.0022}  & 0.74 \\
J$0830.5-4451$ & \multicolumn{1}{c}{0.041} & \multicolumn{1}{c}{0.30}      & 0.55 \\
J$0832.2-4322$c & \multicolumn{1}{c}{0.23}  & \multicolumn{1}{c}{0.46}     & 0.26 \\
J$0833.3-4342$c & \multicolumn{1}{c}{0.41}  & \multicolumn{1}{c}{0.56}   & 0.36\\
J$0833.8-4400$  & \multicolumn{1}{c}{0.018}  & \multicolumn{1}{c}{0.083}    & 0.46 \\
J$0837.8-4048$c & \multicolumn{1}{c}{0.013} & \multicolumn{1}{c}{0.089}     & 0.64 \\
J$0840.5-4122$c & \multicolumn{1}{c}{0.005} & \multicolumn{1}{c}{0.24}    & 0.70 \\
\textbf{J$\boldsymbol{0843.9-4224}$c}    & \multicolumn{1}{c}{$5\times 10^{-5}$}    & \multicolumn{1}{c}{0.055}  & \textbf{0.80}\\
J$0844.1-4330$ & \multicolumn{1}{c}{0.002} & \multicolumn{1}{c}{0.07}   & 0.70 \\
J$0847.8-4138$ & \multicolumn{1}{c}{0.0002} & \multicolumn{1}{c}{0.04}  & 0.76 \\
J$0848.2-4527$ & \multicolumn{1}{c}{0.59}  & \multicolumn{1}{c}{0.11}   & 0.12  \\
J$0848.8-4328$ & \multicolumn{1}{c}{0.0013} & \multicolumn{1}{c}{0.055}   & 0.77 \\
J$0849.2-4410$c & \multicolumn{1}{c}{0.024} & \multicolumn{1}{c}{0.044}    & 0.34 \\
J$0850.3-4448$ & \multicolumn{1}{c}{0.0064} & \multicolumn{1}{c}{0.095}   & 0.66 \\
J$0850.8-4239$ & \multicolumn{1}{c}{$8\times 10^{-5}$}    & \multicolumn{1}{c}{0.072}  & 0.75 \\
J$0851.2-4737$ & \multicolumn{1}{c}{0.018}  & \multicolumn{1}{c}{0.013}  & 0.18 \\
J$0853.1-4407$ & \multicolumn{1}{c}{0.0025}   & \multicolumn{1}{c}{0.091} & 0.76 \\
J$0853.2-4218$c & \multicolumn{1}{c}{0.0006} & \multicolumn{1}{c}{0.038}  & 0.73 \\
\textbf{J}$\boldsymbol{0853.6-4306}$    & \multicolumn{1}{c}{$1\times 10^{-5}$}    & \multicolumn{1}{c}{0.02} & \textbf{0.85}\\
\textbf{J}$\boldsymbol{0854.8-4504}$    & \multicolumn{1}{c}{$2\times 10^{-8}$}    & \multicolumn{1}{c}{$1\times 10^{-6}$}   & \textbf{0.97}\\
J$0856.0-4724$c & \multicolumn{1}{c}{0.19}  & \multicolumn{1}{c}{0.046} & 0.095  \\
J$0857.0-4353$c & \multicolumn{1}{c}{0.033}  & \multicolumn{1}{c}{0.40}  & 0.59 \\
J$0858.4-4615$c & \multicolumn{1}{c}{0.0006}  & \multicolumn{1}{c}{0.06}  & 0.72 \\
J$0859.8-4530$c & \multicolumn{1}{c}{0.0002} & \multicolumn{1}{c}{0.005}  & 0.54 \\
\textbf{J$\boldsymbol{0900.1-4402}$c}    & \multicolumn{1}{c}{$8\times 10^{-6}$}    & \multicolumn{1}{c}{0.01}  & \textbf{0.82}\\
J$0902.8-4633$  & \multicolumn{1}{c}{$6\times 10^{-5}$}   & \multicolumn{1}{c}{0.006} & 0.61 \\ \hline
\end{tabular}
\label{tab1a}}
\tablefoot{Sources that were classified as either AGN or PSR (in our case only pulsars), and their corresponding probabilities, are indicated by bold text.}
\end{table}

Table \ref{tab1b} shows the classification probabilities for the original 4FGL sources in the direction of the Vela molecular ridge (none of which were classified as AGN or pulsar) which are possibly associated to star-forming regions. Section \ref{latsources} explains the model that we adopted for this region.

\begin{table}[ht]
\caption{Classification of sources seen along the Vela molecular ridge}
\resizebox{1\columnwidth}{!}{
\begin{tabular}{lccccc}
\hline\hline
\multicolumn{1}{c}{\textbf{Source}} &
  \multicolumn{2}{c}{\textbf{ANN}} &
  \multicolumn{1}{c}{\textbf{SVM ensemble}} \\ 
& 
  \multicolumn{1}{c}{\textbf{$P(\mbox{\texttt{BLL}})$}} &
  \multicolumn{1}{c}{\textbf{$P(\mbox{\texttt{FSRQ}})$}} &
  \textbf{  $P(\mbox{\texttt{PSR}})$  } \\ \hline
J$0844.9-4117$ & \multicolumn{1}{c}{0.0007} & \multicolumn{1}{c}{0.03}   & 0.72 \\
J$0846.6-4747$ & \multicolumn{1}{c}{0.46}  & \multicolumn{1}{c}{0.43}  & 0.1\\
J$0854.9-4426$ & \multicolumn{1}{c}{0.39}  & \multicolumn{1}{c}{0.34}   & 0.1\\
J$0857.7-4256$c & \multicolumn{1}{c}{0.001} & \multicolumn{1}{c}{0.014}   & 0.58 \\
J$0859.2-4729$  & \multicolumn{1}{c}{0.001} & \multicolumn{1}{c}{0.02}  & 0.72 \\
J$0859.3-4342$  & \multicolumn{1}{c}{0.001} & \multicolumn{1}{c}{0.04}  & 0.75 \\
J$0900.2-4608$  & \multicolumn{1}{c}{0.004} & \multicolumn{1}{c}{0.012}  & 0.55  \\
J$0900.5-4434$c & \multicolumn{1}{c}{$4\times 10^{-5}$}    & \multicolumn{1}{c}{0.03} & 0.69 \\
J$0901.1-4456$c & \multicolumn{1}{c}{0.0002}    & \multicolumn{1}{c}{0.0056}   & 0.75 \\ \hline
\end{tabular}
\label{tab1b}}
\tablefoot{These sources were either kept in the models or replaced by geometrical templates as explained in Section \ref{latsources}.}
\end{table}

It is interesting that we found that the majority of the sources do not belong to either of the two most common classes of point-like LAT sources. We also note that the source 4FGL~J$0830.5-4451$ is located within the $0.37^\circ$-radius source found by \cite{Lange_2025} to the NW of Vela X, and very close to 4FGL~J$0828.4-4444$. We classified both of these 4FGL sources as not belonging to the pulsar nor the AGN classes, and thus they could be related to the PWN or the SNR, as speculated by these authors.

\begin{figure}[h!]
\centering
\includegraphics[width=9cm, height=8.5cm]{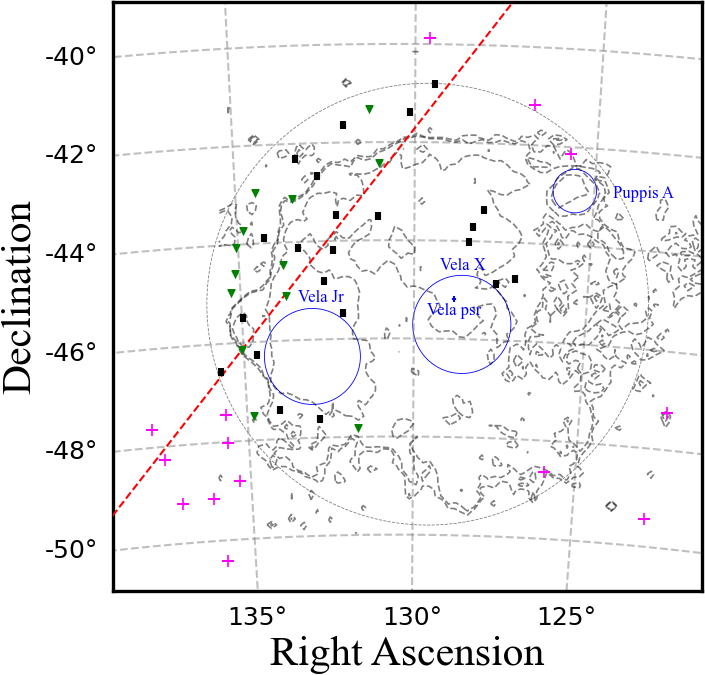}
    \caption{Locations of all the 4FGL sources in the region of Vela, whose contours from a ROSAT observation are shown \citep{1999A&A...349..389V}. The extended sources are labeled (blue circles), as well as the Vela pulsar. The squares represent the sources that we removed from the model (the sources in Table \ref{tab1a} except for the sources classified as pulsars) while the green triangles are the sources that were kept in the model (the sources classified as pulsars in Table \ref{tab1a} and those that might be associated to the Vela molecular ridge, i.e., the sources in Table \ref{tab1b}). The sources that we kept in the model because they have known associations in the 4FGL catalog or lie outside our chosen region around Vela (large dashed circle) are represented by magenta + symbols. The red (dashed) line represents the Galactic plane.}
    \label{fig1}
\end{figure}

A map showing all the LAT sources with the results of our classification and the Vela X-ray shell can be seen in Fig. \ref{fig1}.

\subsection{Extension of the GeV emission}\label{extension}
To improve the background model outside the selected region in Fig. \ref{fig1}, we applied the source-finding algorithm in \texttt{fermipy} requiring a TS $> 16$ and a separation of $0.2^\circ$ from known sources, followed by a fit of the spectral normalizations of sources located within $10^\circ$ of the ROI centre, the spectral parameters of the sources described in Section \ref{latsources}, and the normalizations of the diffuse components. We removed the 22 sources that were classified as \texttt{nonAGN} and \texttt{nonPSR} from the model (see Table \ref{tab1a}). In order to model the residual emission we adopted a simple morphology given by a uniform disk, whose location and size were varied to maximize its TS. During this procedure, we kept free the spectral parameters of all the sources located within $5^\circ$ of the ROI centre. We used a simple power-law spectral function for the disk, $dN/dE = N_0 (E/E_0)^{-\Gamma}$, with free normalization and spectral index ($N_0$ and $\Gamma$). $E_0$ is a fixed scale that we set at the pivot energy, that is the energy for which the propagated uncertainty in $dN/dE$ is minimized. The resulting centre location and radius of the disk are (RA, Dec) $=(131.08\pm 0.03\pm 0.4\,^\circ, -44.32\pm0.05^{+0}_{-1.6}\,^\circ)$, $r=3.24\pm0.03^{+0.6}_{-0}\,^\circ$ (the first error is statistical while the second systematic), and $E_0=2242$~MeV. We estimated these systematic errors by repeating the analysis for each of the 8 alternative models for the diffuse emission developed by \cite{2016ApJS..224....8A}. In all these alternative models we consistently found a larger emission region that is displaced to the south with respect to the disk found using the standard diffuse emission model, and we thus report asymmetric errors for Dec and $r$.

After adding the extended source and again fitting the spectral normalizations for all sources found within $10^\circ$ from the ROI centre and the spectral indices of the sources located within $5^\circ$, we found a TS value for the disk of $2051.5$, corresponding to a high detection significance of $45\sigma$. The resulting spectral index is $\Gamma \sim 2.5$. The emission from the pulsar was effectively suppressed as the corresponding source, 4FGL~J$0835.3-4510$, shows a TS of 84. For comparison we performed an analysis without pulsar gating and obtained a value of the order of $10^6$ for the TS of the pulsar.

\begin{figure}[h!]
\centering
\includegraphics[width=9.5cm]{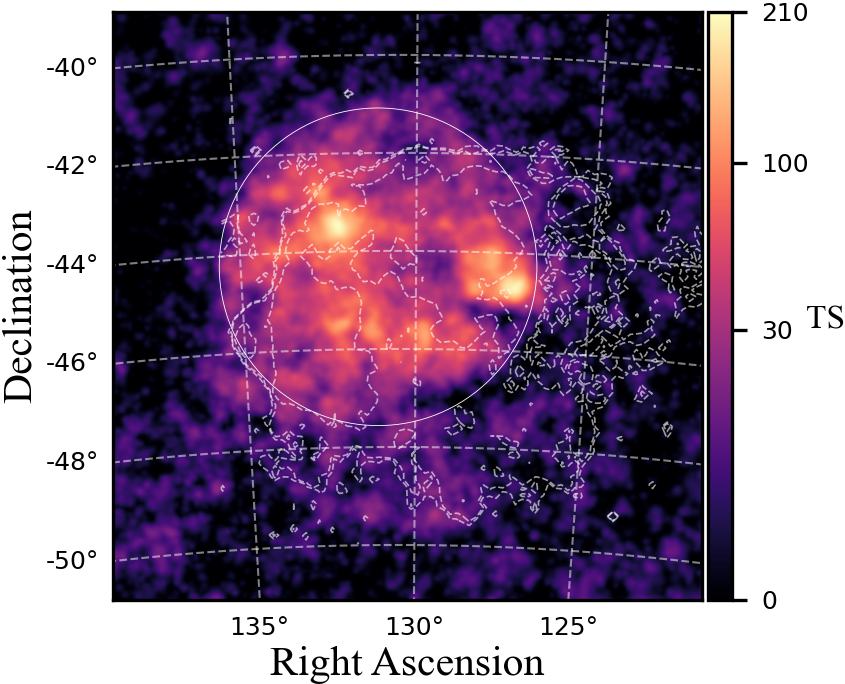}
\includegraphics[width=9.5cm]{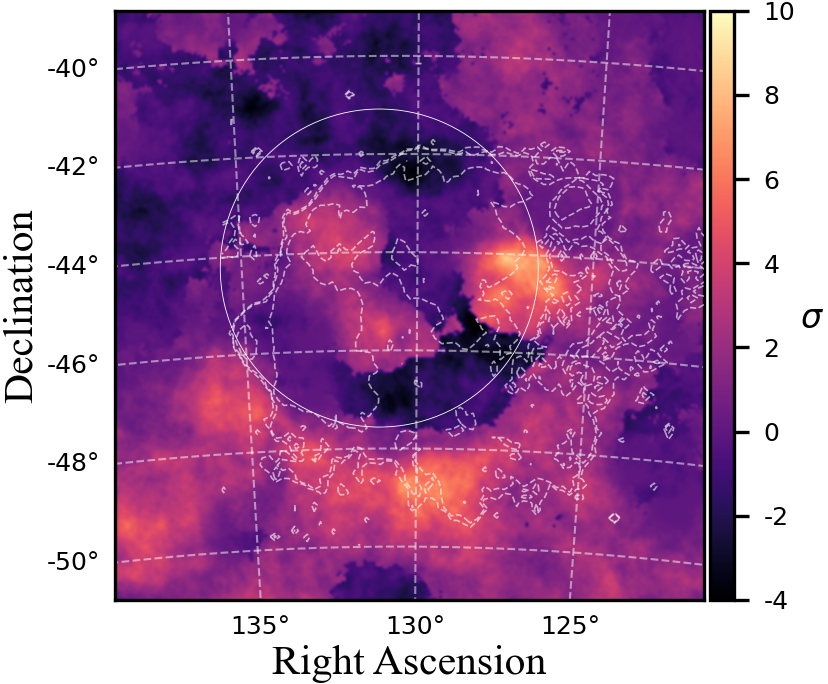}
\caption{\textit{Top.} TS map of the Vela region in the energy range $1-100$~GeV (note the scale is not linear). \textit{Bottom.} Residual PS map (see text). For both maps the contours are the same as in Fig. \ref{fig1} and the large circle represents the disk found in this work.}
\label{fig2}
\end{figure}

In order to visualize the emission we removed the disk from the model and obtained a TS map by fitting the normalization of a test point source at every pixel location assuming a simple power-law spectrum with an index of $2.5$. The result is shown in Fig. \ref{fig2}. To assess the quality of the fit we also calculated a $p-$value statistic map (PS, a data-model deviation estimator) as defined by \cite{2021A&A...656A..81B}. We optimized the map for residuals with a spatial scale of $1^\circ$, and it is shown in Fig. \ref{fig2} in units of $\sigma$. Positive deviations are seen within the shell of the SNR in the southern and western portions, indicating that the uniform disk model adopted in this work is only an approximation to the morphology of the emission. We adopted this model for simplicity and left improvements for future work.

We calculated the Akaike information criterion \citep[AIC,][]{1974ITAC...19..716A} to compare the quality of the model with the extended source (AIC$\mbox{\tiny ext}$) and the original model with the 4FGL point sources (AIC$\mbox{\tiny ps}$). We fit the spectral parameters of the point sources and also considered the location, extension and spectral parameters of the extended source as free in the respective models. The resulting difference is AIC$\mbox{\tiny ps}-$AIC$\mbox{\tiny ext} = 192$, indicating that the model with the extended source is highly preferred over the alternative using the point sources.

\subsection{Gamma-ray spectrum}\label{spectrum}
We repeated the analysis in the entire energy range $0.1-100$~GeV starting with the optimized model found in the previous section and applied again the source-finding algorithm to improve the background as in Section \ref{extension}. We fit the spectral normalizations of the sources found within $10^\circ$ and all the spectral parameters of sources within $5^\circ$ of the RoI centre. When fitting a log-parabola for our disk of the form $dN/dE = N_0 (E/E_0)^{-\alpha-\beta \,\mbox{\small ln}\,(E/E_0)}$ (with $N_0$, $\alpha$ and $\beta$ as free parameters and $E_0=1$~GeV a fixed scale), the TS improves by $\sim 1400$ with respect to the fit using a simple power-law. The source spectrum is significantly curved and the resulting parameters are $N_0=(7.76 \pm 0.006 \pm 4.0)\times 10^{-11}$~MeV$^{-1}$~cm$^{-2}$~s$^{-1}$, $\alpha=2.15\pm 0.0007 \pm 0.04$ and $\beta=0.23 \pm 0.0004 \pm 0.08$ (the first error is statistical while the second systematic). This can be seen in the spectral energy distribution ($E^2\frac{dN}{dE}$, SED) (see plots below). We divided the data in 11 energy bins to estimate the flux values in each bin. For each energy bin, we fit the spectral parameters of the sources with TS $>100$ and set the parameter \texttt{cov\_scale=5} to minimize overfitting and constrain the large number of free parameters with priors taken from the global fit. The resulting fluxes are shown in Table \ref{table2}. We calculated flux upper limits at the 95\% confidence level if the TS of the source is below 4 in a bin.

The source 4FGL~J$0843.9-4224$c, which we classified as a pulsar candidate in Section \ref{sourceclassification}, is not significantly detected with TS$\sim 6$ and therefore we removed it from the model. Regarding the other sources that we classified as pulsar candidates, we obtained TS values of 568 (4FGL~J$0854.8-4504$), 233 (4FGL~J$0853.6-4306$) and 138 (4FGL~J$0900.1-4402$c).

\begin{table}[h!]
\caption{SED fluxes}
\centering
\begin{tabular}{lc}
\hline\hline
Energy range (GeV) & $E^2\frac{dN}{dE}$~($10^{-11}$~erg~cm$^{-2}$~s$^{-1}$)  \\
\hline
$0.1-0.17$  &  $ 6.3^*$\\
$0.17-0.30$  &  $ 10.2\pm 0.40 \pm 6.0$\\
$0.30-0.52$  &  $ 11.8\pm 0.33 \pm 6.1$\\
$0.52-0.91$  &  $11.7 \pm 0.42 \pm 6.2$\\
$0.91-1.6$  & $ 10.6\pm 0.36 \pm 5.4$\\
$1.6-2.8$  &  $ 8.59\pm 0.42 \pm 4.3$\\
$2.8-4.8$  & $ 5.94\pm 0.55 \pm 3.1$\\
$4.8-8.3$  & $ 3.68\pm 0.45 \pm 1.4$\\
$8.3-14.4$  &  $ 3.57\pm 0.59 \pm 1.0$\\
$14.4-37.9$  &  $1.4^*$\\
$37.9-100$  &  $ 1.5^*$\\
\hline
\end{tabular}
\tablefoot{The first error on the fluxes is statistical and the second systematic. Flux upper limits are indicated by $^*$.}
\label{table2}
\end{table}

The integrated energy flux in the $1-100$~GeV range is $\sim 1.9 \times 10^{-10}$~erg~cm$^{-2}$~s$^{-1}$, corresponding to a luminosity of $1.9\times 10^{33}$~erg~s$^{-1}$. For comparison, we estimated the GeV luminosity of Vela X as $\sim 1.4\times 10^{33}$~erg~s$^{-1}$ in the same energy range.

In order to calculate the systematic errors we first considered uncertainties in the LAT's effective area. We propagated the corresponding uncertainty in the spectral parameters by repeating the analysis using a set of bracketing response functions as in \cite{2012ApJS..203....4A}. We also considered the effect of the choice of the 8 alternative models for the diffuse emission \citep{2016ApJS..224....8A} as explained earlier. We repeated the analysis for each alternative diffuse emission model and obtained the corresponding SED fluxes. We added the uncertainties of the two effects described in quadrature and the results are included in Table \ref{table2}. Using the nominal Galactic diffuse emission model we did not detect the source in the last energy bin, $37.9-100$~GeV. Using the alternative diffuse emission models the source was not detected in the energy bin $14.4-37.9$~GeV, as well as in the first bin, $0.1-0.17$~GeV, for three out of the 8 alternative models. We conservatively report upper limits for these three bins obtained with the likelihood profile using the nominal Galactic background model.

As pointed out by \cite{2016ApJS..224....8A} this approach yields a limited estimate of the systematic uncertainty rather than a full determination. Even though there might be additional systematic effects due to the diffuse emission modeling that we did not account for, we believe that the significant detection of GeV photons across the shell of the SNR is strong evidence for its association with this object. We hereafter assume that the gamma rays are produced by Vela and we explore several models for their origin in the following section.

\section{Discussion}\label{discussion}
\subsection{SNR emission}
A uniform distribution of relativistic particles has been proposed to exist in the interior of the Vela SNR \citep{2014A&A...561A.139S}. We used a one-zone model to fit the nonthermal emission from Vela with the \texttt{naima} package \citep{2015ICRC...34..922Z} in two different scenarios. In the leptonic scenario the emission is caused by relativistic electrons. They interact with a magnetic field producing synchrotron emission, and inverse Compton (IC) scatter low energy photons to produce gamma rays. We took the synchrotron fluxes from the northern shell of the SNR \citep[Vela Y, from][]{2001A&A...372..636A} as an approximation to the radio spectrum. We used the two photon fields from \cite{2018A&A...617A..78T} to approximate the local interstellar radiation, given by greybody spectra with temperatures of 30~K (FIR) and 3000~K (NIR), and energy densities of 0.2~eV~cm$^{-3}$ and 0.3~eV~cm$^{-3}$, respectively, as well as the cosmic microwave background (CMB). The IC calculation is adopted from \cite{2014ApJ...783..100K} and the synchrotron emissivity is taken from \cite{2010PhRvD..82d3002A}. We also considered the average magnetic field as a free parameter. In the hadronic scenario, the gamma rays are produced in inelastic scatterings of cosmic ray protons off ambient protons. The parametrization of the gamma-ray production cross section from \cite{2014PhRvD..90l3014K} is implemented in the calculation.

To fit the SED fluxes, we used a differential particle distribution that follows a power-law in momentum, $p$, of the form $\frac{dn}{dp} \propto p^{-s}$, with $s$ the spectral index. We also tested a power-law with an exponential cutoff, $\frac{dn}{dp} \propto p^{-s}\,\mbox{e}^{-p/p_c}$, having an additional free parameter, $p_c$. For the GeV data points we summed the statistical and systematic errors in quadrature to obtain the total error. We compared the fit results using the Bayesian information criterion (BIC) implemented within \texttt{naima}. The model with the lower BIC is favored.

\begin{figure}[h!]
\centering
\includegraphics[width=9cm]{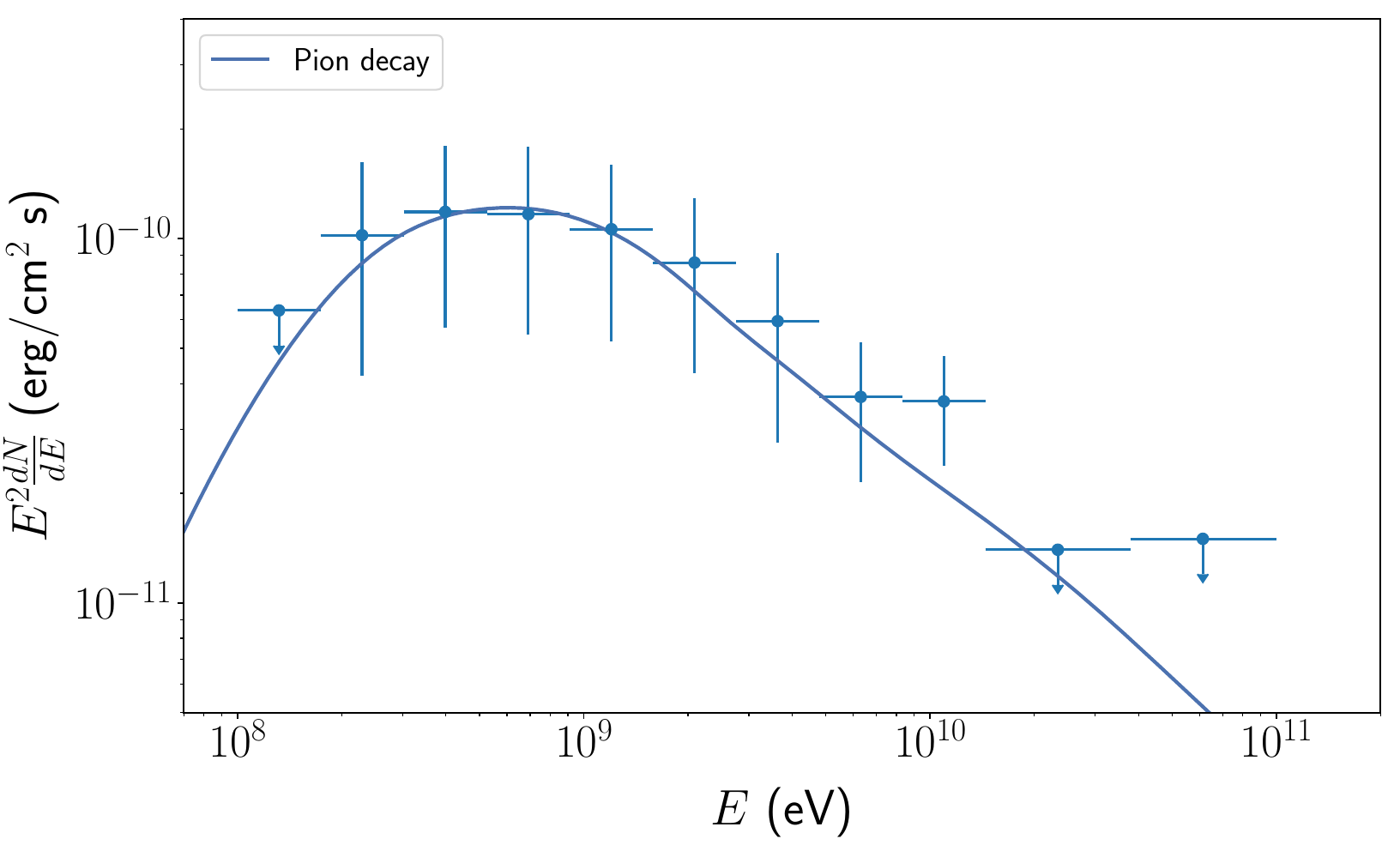}
\includegraphics[width=9cm]{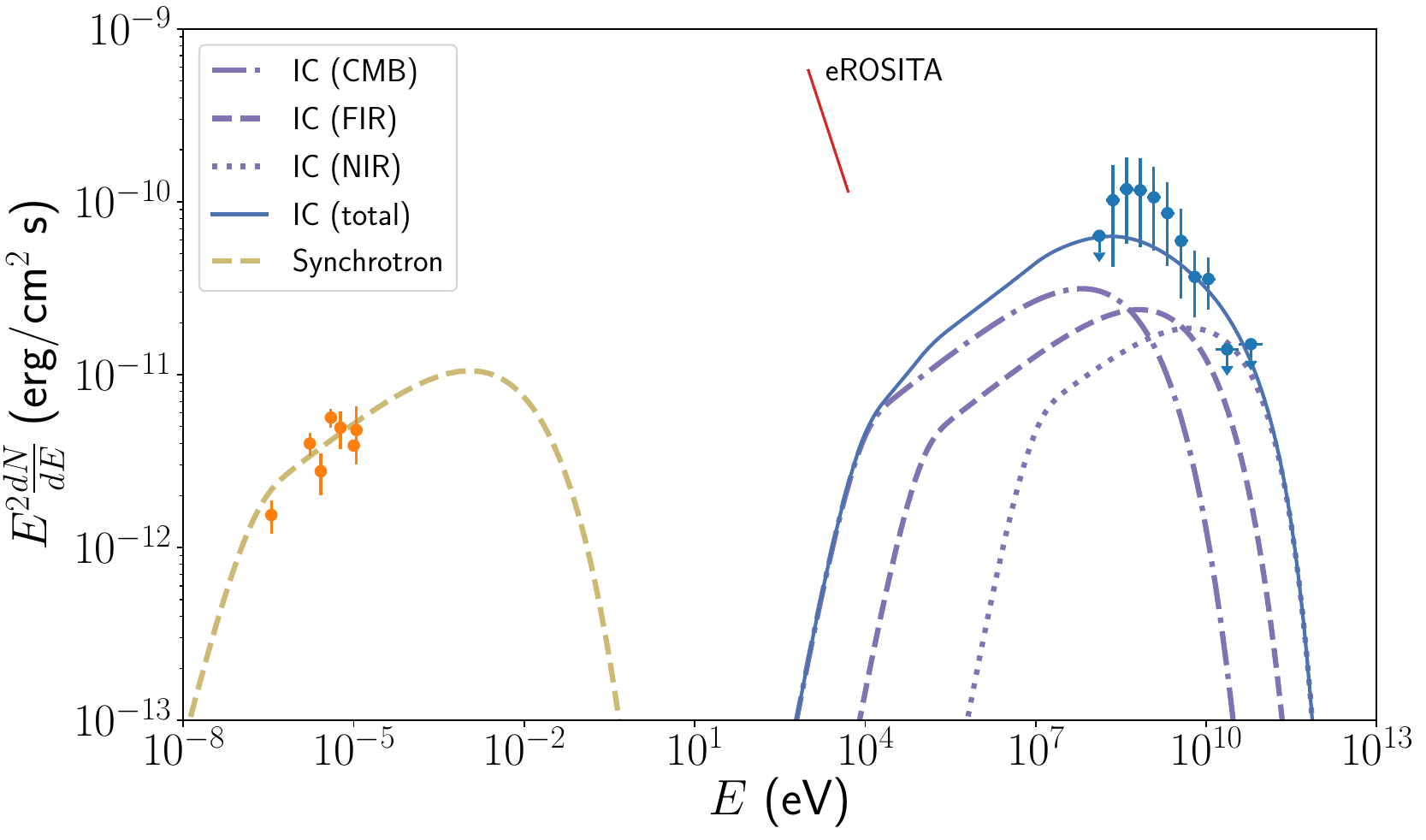}
\caption{\textit{Top}: Gamma-ray fluxes and the hadronic model. The hadronic distribution used is a simple power-law in momentum. \textit{Bottom}: leptonic model for the radio and gamma-ray emission from Vela for a particle momentum distribution which is a power-law with an exponential cutoff (the nonthermal X-ray flux measured by eROSITA is shown for comparison).}
\label{fig3}
\end{figure}

In the hadronic scenario, the power-law particle distribution gives an adequate fit (BIC=5.9) and the cutoff is not required (BIC=7.9). The data and model are shown in Fig. \ref{fig3}. The spectral index is $s=2.97^{+0.14}_{-0.16}$ and the total energy content in the hadrons is $(3.5\pm 1.0)\times 10^{49} \times \left(\frac{n}{\mbox{\tiny 1~cm$^{-3}$}} \right)^{-1}$~erg, where $n$ is the average number density of the ambient material. This energy amount could be provided by an SNR shock with a typical kinetic energy of order $10^{51}$~erg.

In the leptonic scenario, it is not possible to fit simultaneously the radio and GeV fluxes with a simple power-law for the particle distribution because it considerably overpredicts the GeV fluxes. Thus we used a power-law with exponential cutoff. The resulting spectral index and cutoff momentum are $s=2.47_{-0.21}^{+0.12}$ and $p_c = 224_{-96}^{+109}$~GeV/$c$ (with $c$ the speed of light), respectively. The magnetic field is $1.9_{-1.0}^{+2.0}$~$\mu$G and the total particle energy content is $(5.6_{-3.2}^{+5.2})\times 10^{48}$~erg (integrated for particles with energies above 1~GeV). Since it is possible that the radio and GeV fluxes are produced by different leptonic populations, we performed fits to the GeV fluxes only, resulting in a BIC value of 15.1, which indicates that the hadronic model is highly preferred compared to the leptonic fit.

In the plot with the leptonic fit in Fig. \ref{fig3} the approximate nonthermal flux of the X-ray nebula discovered by eROSITA \citep[taken from][and thought to be produced by the pulsar]{2023A&A...676A..68M} is shown for comparison in the $1-5$~keV energy range. We estimated this nonthermal X-ray SED from their reported integrated flux assuming a spectrum which follows a simple power-law with an index of 3 \citep{2023A&A...676A..68M}. However, the spectral index of the nebula varies with location and thus our estimate is only an approximation to the spectrum.

The estimated gamma-ray luminosity in the range $1-100$~GeV of $\sim 3\times 10^{33}$~erg is relatively low compared to those of SNRs interacting with molecular clouds \citep{2016ApJS..224....8A}, but very similar to the luminosity of the Cygnus Loop \citep{2011ApJ...741...44K} and of other relatively faint SNRs detected by the LAT \citep[e.g.,][]{2013MNRAS.434.2202A,2024A&A...684A.150B,2024A&A...691A.225A}. As shown in Fig. \ref{fig2}, the GeV emission is prominent to the NE of the SNR. This is the region where the ambient density is higher and it could be interpreted as supporting the hadronic scenario. \cite{1998AJ....116..813D} showed evidence for the presence of an HI shell in the border of the Vela SNR, suggesting an association with the remnant. This emission seems more prominent in the N and NE of Vela, thus the GeV emission seen in the NE outside the shell of the SNR could be produced by cosmic rays escaping from the object and reaching any atomic gas present. However, several star-forming regions are also seen in the direction outside the NE shell of the SNR, located along the Vela molecular ridge, a cloud complex found farther than the Vela SNR. More detailed work is needed to disentangle any contributions to the GeV emission from the SNR and these star-forming regions.

\subsection{PWN emission}
\cite{2023A&A...676A..68M} recently discovered an X-ray synchrotron nebula that extends for several degrees beyond the known TeV emission region associated with the PWN. High-energy particles could have escaped from Vela X in the past and produce the GeV emission extending through the Vela SNR. From the point of view of the energetics, the total spin-down energy injected by the pulsar over its lifetime is more than sufficient to account for a particle content of $\sim 10^{48}$~erg \citep{2010ApJ...713..146A}, as required by the simple leptonic estimate in the previous section. The emission found in this work has a GeV spectrum that is softer than that reported for Vela X in the catalog and a comparable luminosity. First of all, the one-zone  leptonic scenario presented in the previous section provides a relatively poor fit to the data. Furthermore, \cite{2011ApJ...743L...7H} have shown that the particles escaping from Vela X should produce GeV emission with a spectrum that is much harder, and with considerably higher luminosity, than what we found in the Vela SNR. Even though our leptonic fit in the previous section is poor, it implies a particle cutoff energy of $\sim 300$~GeV, while the highest energies of the escaping particles in their model are in the TeV range. Although a high enough magnetic field could produce the spectral turnover, a magnetic field value of only a few $\mu$G is required to explain the extended X-ray nebula \citep{2023A&A...676A..68M}. Detailed modeling is needed to explain the role, if any, that Vela's pulsar could have in producing the GeV emission. On the other hand, the GeV emission found in this work cannot be produced by the same electrons responsible for the extended X-ray nebula. The typical energies of electrons that emit X-ray synchrotron emission in a magnetic field of a few $\mu$G are well into the TeV regime. A considerably extended TeV counterpart of the X-ray nebula will likely be revealed with future observations \citep[for example by the SWGO experiment,][]{2025arXiv250601786S}.

\section{Conclusions}
GeV emission is seen across the shell of the Vela SNR after subtracting the contributions from the sources in the 4FGL LAT catalog in the region. Besides this emission, a cluster of unidentified ``point sources'' located mainly across the NE shell of the Vela SNR is found in the catalog. This may indicate that the point sources are actually part of the SNR. We applied machine learning classification schemes to this set of sources and found that the majority do not share spectral features with those of known populations of GeV-emitting AGNs and pulsars, the most common types of point sources in the LAT catalogs. We thus proposed that these ``point-like sources'' are part of the extended SNR. We modeled the emission with a simple geometrical template whose best-fit size is close to that of the Vela SNR, although it is displaced to the NE. The emission shows a high detection significance ($\sim 45\sigma$, $1-100$~GeV). The GeV luminosity is similar to that of other SNRs. The spectrum of the gamma-ray emission and its spatial coincidence with regions of enhanced ambient density point to a hadronic origin, while a one-zone leptonic model provides a worst fit to the GeV fluxes. A contribution from the PWN Vela X is problematic due to the soft GeV spectrum of the observed emission, its relatively low luminosity compared to predictions from the population of leptons escaping from Vela X and the difficulties in fitting the spectral shape with a leptonic model. More detailed models for the evolution of the system should be carried out in order to assess the role that Vela X could have played in producing the GeV photons. Future studies to characterize the gamma-ray emission on smaller angular scales throughout Vela should be possible. In addition, the source classification scheme carried out in this work could constitute a curated analysis-ready corpus for \textit{Fermi}-LAT sources in general, which may prove valuable for reproducibility of this work and for future independent studies of any sky region with the LAT.

\begin{acknowledgements}
We are grateful to the anonymous referee for providing a valuable review which helped improve this work. We also thank M. Kerr for the updated ephemeris of the Vela pulsar, and funding from Universidad de Costa Rica under grant number C4228.
\end{acknowledgements}

\bibliographystyle{aa}
\bibliography{vela.bib}

\end{document}